\documentclass{mn2e}

\usepackage{color}
\usepackage{graphicx}
\begin{document}

\title[Gradients in spheroids]{Abundance gradient slopes versus mass  
in spheroids: predictions by \emph{monolithic} models}
%\titlerunning{Gradients in spheroids}

\author[A. Pipino et al.]{Antonio Pipino$^{1,2}$, Annibale D'Ercole$^3$, Cristina Chiappini$^{4,5}$ and Francesca Matteucci$^{1,5}$\\%}
%\authorrunning{A. Pipino et al.}
%\institute{
$^1$ Dipartimento di Fisica, sez.di Astronomia, Universit\`a di Trieste, Via G.B.  
Tiepolo 11, 34100 Trieste, Italy\\
$^2$ Department of Physics \& Astronomy, University of California, Los Angeles,
430 Portola Plaza, Box 951547, Los Angeles, CA 90095 ,USA \\
$^3$ INAF-Osservatorio Astronomico di Bologna, via Ranzani 1, 40127  
Bologna, Italy\\
$^4$Observatoire de Gen\`eve, Universit\'e de Gen\`eve,
51 Chemin de Mailletes, CH1290, Sauverny, Switzerland\\
$^5$ INAF, Osservatorio Astronomico di Trieste, Via G.B. Tiepolo 11,  
34100 Trieste, Italy}

\date{Accepted 2010 May 10.  Received 2010 April 26; in original form 2009 June 18}
%\begin{document}
\maketitle

%\maketitle

\begin{abstract}
%We aim at showing that elliptical galaxies formed through dissipative  
%collapse
%exhibit a range of radial gradients in the stellar metallicity  
%comparable to
%the observed one.
{ We investigate whether it is possible to explain the wide range of observed gradients in early type galaxies in the framework of monolithic models}.
%The main scope is to understand whether a correlation
%of the gradient slope with the galactic mass exists.
To do so, we extend the set of hydrodynamical simulations by Pipino  
et al. (2008a)
by including low-mass ellipticals and spiral (true) bulges.
These models
satisfy the mass-metallicity and the mass-[$\alpha$/Fe] relations.
The typical metallicity
gradients predicted by our models
have a slope of -0.3 dex per decade variation in radius,
consistent with the mean values of several observational samples.  
However, we
also find a few quite massive galaxies in which this slope is -0.5  
dex per
decade, in agreement with some recent data.
In particular, we find a mild dependence from the mass tracers when  
we transform
the stellar abundance gradients into radial variations of the $Mg_2$  
line-strength
index, but not in the åÊ$Mg_b$. We conclude that, rather than a mass- 
slope relation,
is more appropriate to speak of an increase in the scatter of the  
gradient slope {with the galactic mass}.
%at a fixed mass.
We can explain such a behaviour with different efficiencies of star  
formation
in the framework of the revised monolithic
formation scenario, 
hence the scatter in the observed gradients should not be used as an evidence of the need of mergers.
%rather than by splitting galaxies into several  
%categories according
%to their formation mechanism; namely truly monolithic ones versus  
%merger outcomes, {as recently suggested in the  
%literature
Indeed, model galaxies that exhibit the steepest gradient slopes are preferentially those with the
highest star formation efficiency at that given mass.
\end{abstract}

\maketitle

%\keywords{}

\section{Introduction}

Negative metallicity radial gradients in the stellar populations are  
a common
feature in spheroids (e.g. Carollo et al. 1993, Davies et al. 1993)
and must be predicted by every theory for the formation of elliptical  
galaxies.
A possible fingerprint of a given galaxy formation scenario
might be the (lack of) correlation between gradient properties (e.g.  
the slope)
and either global galactic properties (namely mass, stellar velocity  
dispersion $\sigma$,
total magnitude) or central ones (e.g. central metallicity or [$< 
\alpha$/Fe$>$]). From the theoretical point of view, in fact, steep metallicity gradients are  
expected from classical dissipative collapse models (e.g. Larson 1974, Chiosi \& Carraro, 2002)
and their (\emph{revised}) up-to-date versions  which start from semi- 
cosmological initial conditions (e.g. Kawata 2001, Kobayashi, 2004).
The abundance gradient arises because the stars form everywhere in a  
collapsing cloud and then remain in orbit
with a little inward motion,\footnote{Stars will spend most of their
time near the apocentre of their orbit.}
whereas the gas sinks further in because of dissipation.
This sinking gas contains the new metals ejected by evolving stars so  
that an abundance gradient develops in the gas.
As stars continue to form their composition reflect the gaseous
abundance gradient.
{ The original dissipative models} predict a steepening of the gradient as the galactic  
mass increases, mainly
because the \emph{central} metallicity is quickly increasing with mass 
\footnote{The fit of the mass-metallicity
relation, namely the increase of the mean metal content in the stars  
as a function of galactic mass (O'Connel, 1976), was the main success of these {}
original models.}, whereas the \emph{global} one has a milder variation (Carlberg 1984).
{ At the same time they} predict metallicity gradient as steep as -0.5 dex  
per decade variation in radius.
{ On the other hand, the few attempts
to study the gradients in the merger-based models  
hint for very shallow (if any) gradient (Bekki \& Shyoia, 1999),
less steep than the mean observational values and than
the predictions from monolithic collapse models. Moreover, it seems
that dry mergers flatten pre-existing gradients (Di Matteo et al., 2009).
Indeed, when the two scenarios (monolithic collapse and mergers) are considered
as two possible channels working at the same time, the scatter in the predicted gradients for such a population of galaxies seem to
be in agreement with observations (Kobayashi, 2004).}

More recently, observations showed that successful models for elliptical galaxies 
should also reproduce the [$<\alpha$/Fe$>$]-mass relation (Worthey et al. 1992, Thomas et al. 2007)
as well as the observed
%This result seems to be questioned by a handful of more recent observational
%works {which } 
gradients in the [$<\alpha/Fe>$] ratios
(Mehlert et al. 2003, Annibali et al. 2007,
Sanchez-Blazquez et al. 2007, Rawle et al., 2008). {Indeed, these   
observations} show that the slope in the [$<\alpha/Fe>$] gradient has a typical value close to zero
and does not correlate with mass.
%{This suggests that the outside-in formation scenario for ellipticals galaxies is  
%not a sufficient condition for the formation of the abundance ratio
%gradients (see Pipino et al. 2008a, hereafter Paper I)}.

These observations have been
interpreted by Pipino et al. (2008a; Paper I, hereafter) 1D hydrodynamical code 
as the fact that the suggested outside-in mechanism for the formation of the ellipticals is not the only process
responsible for the formation of gradients in the abundance ratios.
Other processes should be considered such as the interplay between
the star formation (SF) timescale and gas flows. While such an interplay
flattens the [$<\alpha/Fe>$] gradient to the value required
by observations, it still enables galaxies
to harbor gradients in [$<$Fe/H$>$] and [$<$Z/H$>$] in agreement with the most
recent observations (see Section 2).
%Here we note that the shallower gradient predicted by this new monolithic
%models with respect to earlier dissipational collapse models as well
%as the lack of correlation between gradient slope and galactic mass
%stem from the fact that it is necessary to produce stars with high [$<\alpha 
%$/Fe$>$] ratios inhabiting the galactic core. 
%Such a constraint, that was not available when the original collapse models,
%turns into a need  
%for short duration of the star formation. Hence the
%metal enrichment process in the centre is faster, leading to smaller  
%age gradients} even in the most massive galaxies. 
Pipino et al. (2008b; Paper II) calibrated such a model
by means of the resolved stellar populations in the Milky Way bulge.
As a matter of fact, spiral \emph{true}\footnote{In the rest
of the paper we will consider only the class of \emph{true}
bulges, (Kormendy \& Kennicutt, 2004).} bulges remarkably follow many  
fundamental constraints for ellipticals
such as the mass-metallicity and the mass-[$<\alpha$/Fe$>$] relations  
(see below), the only difference
being that they might be rejuvenated systems (Thomas \& Davies, 2006).

The aim of this paper is to explore a wider range of cases
by extending the analysis of Paper I to lower masses, including bulges,
and compare them to the latest observational results.
{ In this way we can study the correlation between gradient slopes and  
galactic mass
(if any) in order to understand whether the monolithic galaxy formation scenario is
in agreement with the recent observational evidences}.

In Sec.~\ref{obs} we give a brief overview of the observations regarding
metallicity gradients in ellipticals.
The main characteristic of the model are briefly described in Section~ 
\ref{model}.
We characterize the global properties of our models in Section 4, present our results in Section~\ref{results}, discuss
them in Section 6 and draw our conclusions
in Section~\ref{concl}.

\section{The observational background}
\label{obs}

In general, observations show that the majority of ellipticals has
as typical decrease in metallicity of 0.2-0.3 dex per decade in  
radius (e.g. Carollo et al. 1993, Davies et al. 1993).
However, a large scatter in the gradient slope at a given galactic  
mass is also observed. The exact slope depends on the line-strength index used to infer  
the metallicity.
Below we give a brief historical perspective for what concerns the  
relation
between gradient slope and mass. åÊWe refer the reader to other  
works (e.g. Sanchez-Blazquez et al. 2006) for a %thorough
review about the debate on the observations in the literature.

Indeed, a positive correlation of the metallicity gradient slope with
the galactic mass - namely gradients becoming more negative at higher
galactic masses - {(in agreement with Larson 1974's  
prediction), has been reported by} Carollo et al. (1993), but only
for masses lower than $10^{11}M_{\odot}$. In fact, Carollo et al.  
(1993) found a flattening of
the observed gradients in the most massive galaxies of their sample  
and ascribed this fact to: i) an increase in the importance of  
mergers; or ii) a less
important role of dissipation in the formation of the most massive  
galaxies.
The positive correlation of the slope with the galactic mass was  
later confirmed
by some authors (e.g. Gonzalez \& Gorgas, 1996) over the entire mass  
range and denied by others who either found no statistical
evidence for such a correlation (e.g. Kobayashi \& Arimoto, 1999) or  
a very mild opposite trend
(e.g. Annibali et al., 2007). We notice that { several} of the studied samples
were quite small or not homogeneous (e.g. Kobayashi \& Arimoto, 1999).
In recent years, a positive correlation of gradient slope with mass  
has been
suggested again by Forbes et al. (2005), Sanchez-Blazquez et al  
(2007), for the entire mass range of elliptical galaxies.
Ogando et al. (2005), rather than a clear trend, noticed an  
increasing number of E and S0 galaxies harboring steep $Mg_2$  
gradients with
increasing velocity dispersion.
Interestingly, Spolaor et al. (2009) found {a similar  
result} for massive ellipticals, whereas,
for the first time, detected a clear gradient slope-mass relation at  
low mass end (Fornax and Virgo dwarf).
%Hence, they confirmed the original suggestion by Carollo et al.  
%(1993) of two regimes
%in the gradient slope-galactic mass relation.
{ Spolaor et al's result has been questioned by Koleva et al. (2009a), who do not 
observe any such a trend in another sample of dwarf galaxies in the Fornax cluster.
To date, no one has offered a convincing explanation for the discrepancy
between observational results (unfortunately Koleva et al.'s and Spolaor
et al's samples do not overlap!). One problem, of course, is the small number statistics. Issues related to the reduction and analysis process have been excluded
as a cause for this discrepancy (Koleva et al., 2009b).
Moreover, as we will discuss later in the comparison
between our models and observations, different authors use
different (combinations of) indices to estimate the age, $\alpha$/Fe and
metallicity indices. This is sufficient to make the inferred gradients
appear either stronger or weaker (e.g. Sanchez-Blazquez et al., 2006). 
In addition, they use different SSP libraries and minimization techniques to transform their data into
metallicity (either [Z/H] or [Fe/H]) and ages, thus introducing further issues in the interpretation (see
Pipino et al., 2006 for an extended discussion).}

Bulges have gradients in metallicity (Goudfroij et al. 1999, Proctor  
et al. 2000)
and [$<\alpha/Fe>$] ratios (Jablonka et al. 2007) with the same  
properties
as those in ellipticals. In particular, Jablonka et al. {
(2007)} described the
variation in the gradient slope as a function of mass as a multi-step  
process rather than a smooth transition in gradient amplitude
with velocity dispersion. According to {the latter  
authors}, at
large masses the dispersion among gradients is large but small  
gradients are relatively rare. At smaller masses, instead,
galaxies with very weak gradients appear in larger number.

\section{The model} %%% Top level section head (remove "%" symbol)
\label{model}

We adopted a one-dimensional hydrodynamical model (Frankenstein)  
that follows
{ the
time evolution of the density of mass ($\rho$), momentum ($m$) and
internal energy ($\varepsilon$) of a galaxy, under the assumption of  
spherical
symmetry.  In order to solve the equation of hydrodynamics with a source
term we made use of the code presented in Ciotti et al. (1991), which
is an improved version of the Bedogni \& D'Ercole (1986) Eulerian,
second-order, upwind integration scheme (see their Appendix).
Here we report the gas-{dynamics} equations:
\begin{equation}
{\partial\rho\over\partial t}+\nabla\cdot(\rho {{u}})=
\alpha\rho_* -\Psi,
\end{equation}
\begin{equation}
{\partial\varrho^i\over\partial t}+\nabla\cdot(\varrho^i {{u}})=
\alpha^i\rho_*-\Psi \varrho^i /\rho	,
\end{equation}
\begin{equation}
{\partial {{m}}\over\partial t}+\nabla\cdot({{m}}
{{u}})=\rho{{g}}-(\gamma-1)
\nabla\varepsilon %+ \alpha {\rho_*} {{u}}_*
-\Psi {{u}} ,
\end{equation}
\begin{equation}
{\partial\varepsilon\over\partial t}+\nabla\cdot(\varepsilon {{u}})=
-(\gamma-1)\varepsilon\nabla\cdot{{u}}-L+\alpha\rho_*
\biggl(\epsilon_0+{1\over 2}u^2\biggr)-\Psi \varepsilon /\rho\, .
\end{equation}
\noindent
The parameter $\gamma=5/3$ is the ratio of the specific heats, ${{g}} 
$ and
${{u}}$ are the gravitational acceleration { due to the total mass distribution
(stars and dark halos)} and the fluid
velocity, respectively. The source terms on the r.h.s. of equations
(1)--(4) describe the injection of total mass and energy in the gas due
to the mass return and energy input from the stars.
%${{u}}_*$ is the circular velocity of these stars, and
$\alpha(t)=\alpha_*(t)+\alpha_{\rm SNII}(t)+\alpha_{\rm SNIa}(t)$ is
the sum of the specific mass return rates from low-mass stars and SNe of
both Type II and Ia, respectively. $\epsilon_0 = 3 kT_0/(2\mu m_p)$ is the injection energy per unit mass due to the SN explosions,
and $T_0$ is the injection temperature. % (see, e.g. Lowenstein \& Mathews, 1987).
The positive source term on the right-hand side of the energy equation describes the
heating of the gas by SN blast waves and by the relative motion of the mass-losing
stars and the ISM (kinetic heating). 
$\Psi$ is the astration term due to SF.  $L=n_{\rm e}n_{\rm
p}\Lambda(T,Z)$ is the cooling rate per unit volume, where for the
cooling law, $\Lambda(T,Z)$, we adopt the Sutherland \& Dopita (1993)
curves.  This treatment allows us to implement a self-consistent
dependence of the cooling curve on the metallicity (Z) in the present
code.  We do not allow the gas temperature to drop below $10^4$
K, as the Sutherland \& Dopita (1993) functions are calculated only above this limit.
We are aware that fixing the minimum gas temperature can be a limitation of 
the model, but this is done in order to avoid the complexity of the cooling at lower 
temperatures. 
{ Moreover, as it can be seen from Paper I Figs. 1 and 2, at the time of the occurrence of the winds
(and actually for most of the pre-wind evolution) the majority of  
the models exhibit T $>> 10^4$ K.}

$\varrho^i$ represents the mass density of the $i-th$ element, and
$\alpha^i$ the specific mass return rate for the same element, with
$\sum^N_{i=1} \alpha^i=\alpha$. Eq. (2) represents a
subsystem of four equations that follow the hydrodynamical evolution
of four different ejected elements (namely H, He, O and Fe).  
This set of elements is good enough to characterize our
simulated elliptical åÊgalaxy from the chemical evolution point of  
view.
We divide the grid in 550 zones 10 pc wide in the innermost regions, and
then slightly increasing with a size ratio between adjacent zones
equal to 1.03.
At the same time, however, the size of the simulated box is roughly
a factor of 10 larger than the stellar tidal radius.
This is necessary to avoid possible perturbations at the boundary
affecting the galaxy and because we want to have a surrounding medium  
that acts as a gas reservoir for the models.
We adopted a reflecting boundary condition in the center of the grid
and allowed for an outflow condition in the outermost point.

At every point of the mesh we allow the SF to occur at the following  
rate:
\begin{equation}
\Psi = \nu  \rho = {\epsilon_{SF} \over max(t_{cool},t_{ff})} \rho \,
\label{sfr}
\end{equation}
where $t_{cool}$ and $t_{ff}$ are the \emph{local}
cooling and free-fall timescales, respectively,
and $\epsilon_{SF}$ is a suitable \emph{SF parameter}
that contains all the uncertainties on the timescales of the SF  
process that
cannot be taken into account in the present modelling
 and will be taken as a free parameter in our  
models. { In fact, star formation is an inherently 3D process which cannot be even approximately
simulated by 1D simulations. Moreover, star formation
occurs on small scale, much smaller than any possible mesh  
resolution when the whole galaxy must be covered by the numerical grid.
We recall that the \emph{final efficiency},
namely the fraction of gas that eventually turned into stars, is an  
output of the model.}

We assume that the stars do not move from the gridpoints at which they
have been formed, since we expect that the stars will
spend most of their time close to their apocentre.}

%The Paper I main novelty {was} that
%we follow the chemical evolution of several elements, namely H, He,
%O and Fe. 
%This set of elements is good enough to characterize our
%simulated elliptical åÊgalaxy from the chemical evolution point of  
%view. åÊ
%We refer the reader to Paper I for a comprehensive
%discussion of the adopted code, a rigorous testing of its behaviour
%in relation to PM04 chemical evolution code and for the thorough  
%description
%of the build-up of the gradients for massive spheroids.

\subsection{Chemical evolution}

{ The nucleosynthetic products enter the mass conservation  
equations via several source terms,
according to their stellar origin.
A Salpeter (1955) initial mass function (IMF) constant in time in the
range $0.1-50 M_{\odot}$ is assumed.
We adopted the yields
from Iwamoto et al. (1999, and references therein) for both SNIa and
SNII.  The SNIa rate for a~SSP formed at a given radius is
calculated assuming the single degenerate scenario and the Matteucci
\& Recchi (2001) delay time distribution (DTD). }
%The convolution of
%this DTD with $\Psi$ over the
%galactic volume gives the total SNIa rate, according to Greggio  
%(2005). 
{ These quantities, as well as the evolution of single low and  
intermediate
mass stars, were evaluated by adopting
the stellar lifetimes given by Padovani \& Matteucci (1993).
The solar abundances - { used to present our values in the ``[$<\, >$]''notation} - are taken from Asplund et al. (2005),
unless otherwise stated. { Note that, as far as gradient slope are concerned, the actual solar
scale does not make any difference.}

In order to study the mean properties
of the stellar component in ellipticals, we need average quantities
related to the mean abundance pattern of the stars, which, in turn,
can allow a comparison with the observed integrated spectra.
%To this scope, we recall that, at
%a given radius, both real and model galaxies are made of CSP,
%namely a mixture of several SSP, differing
%in age and chemical composition according to the galactic chemical
%enrichment history, weighted with the SF rate.  On the other hand, the
%line-strength indices are usually tabulated only for SSPs as functions
%of their age, metallicity and (possibly) $\alpha$-enhancement {(see PMC06 for a discussion on this point)}.
{ In particular, we make use of the luminosity-weighted mean stellar abundances.
Following Arimoto \& Yoshii (1987), we have:
\begin{equation}
\rm <O/Fe>_V=\sum_{k,l}n_{k,l} (O/Fe)_l L_{V,k} / \sum_{k,l}n_{k,l} L_{V,k}    \, ,
\label{AY87}
\end{equation}  
where $n_{k,l}$ is the number of stars binned in the interval
centered around $\rm (O/Fe)_l$ with V-band luminosity $\rm L_{V,k}$.
We then take the logarithm and express the quantities in solar units.
Similar equations hold for [$<Fe/H>_V$] and the global metallicity [$<Z/ 
H>_V$]. 
Generally the mass averaged {\rm [Fe/H] and [Z/H]} are slightly larger than the luminosity
averaged ones, except for large galaxies (see Yoshii \& Arimoto, 1987). 
We will present our results in terms of $[<Fe/H>_V]$ and $[<Z/H>_V]$,
because the luminosity-weighted mean is much closer to the actual
observations and might differ from the mass-average, unless otherwise stated.
Therefore we drop the subscript $V$ in the remainder of the paper.}

\subsection{Model classification and initial conditions}

The initial set-up of the new simulations for low-mass ellipticals is  
presented in Table 1,
where the name of the model, the gas density ($\rho_{core,gas}$ )
as well as the initial gas temperature, the star formation  
parameter $\epsilon_{SF}$ and the Dark Matter (DM) halo mass are reported.
In the same table we include also the models already presented in Paper I and the bulges (see below).

{ We recall that in Paper I we defined the following two families  
of models according to the total initial DM and gas
content{: Model M -- $2.2\cdot 10^{12} M_{\odot}$ DM halo  
and $\sim 2 \cdot 10^{11} M_{\odot}$ of gas;
Model L -- $5.7\cdot 10^{12} M_{\odot}$ DM halo and $\sim 6.4\cdot 10^ 
{11} M_{\odot}$ of gas.}
{ The DM potential has been evaluated by assuming a distribution
inversely proportional to the square of the radius at large
distances (Silich \& Tenorio-Tagle 1998).}
These quantities have been chosen to ensure a final ratio between the  
mass of baryons
in stars and the mass of the DM halo of around 0.1.

In order to model ellipticals less luminous than the ones presented  
above, we assume that the galaxy assembly occurs in a 0.3 times smaller and  
0.1 times lighter
DM halo. This guarantees that we model $\sim 0.5-2 \times 10^ 
{10} M_{\odot}$ galaxies
(stellar mass) in $\sim 2 \times 10^{11} M_{\odot}$ Dark Matter haloes.
{ Note, however, that the final mass in
stars is determined by the interplay between the SF efficiency
and the duration of the SF process itself (regulated by infall
and stellar feedback). Therefore we may have two galaxies
with the same DM potential and differing stellar masses because
of the different evolutionary paths.} 
%Models by  Matteucci (1992) and PM04 require
%such a ratio for ellipticals in order to develop a galactic wind.
%Concerning the class of models labelled \emph{Ma} and \emph{La}, we  
%mainly vary the gas temperature and the
%parameter of star formation. 

%For the class of models labelled \emph{Mb} and \emph{Lb} , instead,  
%the initial gas density (as
%reported in Table 1 under the column $\rho_{core,gas}$ ) can be a
%crucial parameter, as well as the gas temperature and $\epsilon_{SF}$.
%Here the values for $\rho_(r,t=0)=\rho_{core,gas}$
%may be lower than in the \emph{a} cases,
%otherwise we would have too much gas in the grid, even
%{ is chosen in order to have the initial gas content in the whole  
%grid not} higher than the typical
%baryon fraction in a high density environment (i.e. 1/5-1/10 as in
%a galaxy cluster, e.g. McCarthy et al. 2007).
}

Concerning the bulges, instead, we assume that they are stellar systems
with mass $\sim 2 \times 10^{10} M_{\odot}$ embedded in a $\sim 100$  
times more massive Dark Matter halo,
since bulges occupy only the central part of their large hosts.
We neglect the presence of a disc, which requires a much longer  
timescale to be built { (e.g. Zoccali et al., 2006 from the observational viewpoint,
Matteucci \& Brocato, 1990, Ballero et al.,2007, from the theoretical one).
Moreover, Sarajedini \& Jablonka (2005) suggest a common scenario for the formation of bulges
that is not linked to the host galaxy formation.
Finally, the observations we are comparing our results to have been derived by accurately
selecting edge-on galaxies. Therefore the contamination from the discs should be minimal.}

To generate different models we mainly vary the gas temperature and the
efficiency of star formation as well as the initial gas density  
distribution.

In particular, the gas can initially be an isothermal sphere (models  
flagged as \emph{IS}) in equilibrium within the galactic
potential well (i.e. due to both
DM and gas). The actual initial temperature is lower than the virial
temperature, in order to induce the gas to collapse.
This is an extreme case in which we let all the gas
be accreted before the SF starts. In other models, instead, the gas has
uniform distribution within the whole computational box (models  
flagged as \emph{flat}).
At variance with the previous models, in this case we let the SF
process start at the same time as the gas accretion.
%For the class of models labelled \emph{Sb}, instead, the initial gas  density (as
%The main input parameters for the new models are given in Table ~\ref {table1},
%crucial parameter,
{ The values for $\rho_{core,gas}$ are set in order not to have too much gas in the grid,
namely higher than  the typical
baryon fraction in high density environment (i.e. 1/5-1/10 as in
galaxy cluster, e.g. McCarthy et al. 07).}

The initial gas temperature ranges from $10^{4-5}$ K (cold-warm gas)
to $10^{6-7}$ K (virialised haloes).
{ This range of temperature is consistent with the typical findings of simulations
of high-redshift galaxy formation. In fact, a common assumption in galaxy formation models has that the gas accreted by a DM halo is shock-heated to the host halo virial temperature ($10^{7}$K), and only
then is able to cool down and feed star formation (e.g. White \& Rees 1978). This scenario justifies the models
with a high initial temperature and \emph{IS} gas profile, with the only difference
that the amount gas reservoir is not regulated by any ``cosmological'' infall history.
Slightly different (high) initial temperatures may be used to regulate/delay the \emph{infall} rate on the actual protogalaxy. 
We note that the gas cools very rapidly, therefore the actual starting value is less
important than, e.g., the chosen $\epsilon_{SF}$.
Recent simulations show that the gas may be accreted through cold filaments (e.g. Dekel \& Birnboim, 2006)
streaming through the shock-heated gas. In this case it will have temperatures of about $10^{4-5}$K.
The majority of the models presented here (\emph{flat} profile and warm temperature) are motivated by these recent results. The 1D nature of our study hampers us to mode
these ``cold'' accretion flows, therefore we simply varied
intial gas density and temperature in order to give a reasonable approximation to this picture.}

In general, { we assume values for the star formation parameter between
0.1 and 10. These values guarantee star formation rates of 10-500 $M_{\odot}$/yr (c.f
Paper I, Fig. 8) in massive galaxies, comparable with the observations of high redshift star
forming objects. A preliminary exploration of the parameter
space returned that smaller values of the star formation parameter give rise to too extended star formation
histories (and hence too low [$\alpha/Fe$] ratios). On the other hand,
higher values would lead to too small (in terms of stellar to total mass ratio) and too 
large (in terms of $R_{eff}$) galaxies - similarly to a what happens when
one adopts a 100\% SN efficiency (cf model MaSN, Paper I). In fact, 
the strong feedback from SNe halts the SF too early by preventing further accretion of gas. 
Such a galaxy would also have a too high [$\alpha$/Fe]. These models have been discarded
during the preliminary analysis that led to Paper I.}
Both the SNIa and SNII efficiency is assumed to be
constant $\epsilon_{SN}=0.1$ (see Paper I, Pipino et al. 2005).

%\par
%\hbox{}
\begin{table*}
\centering
\begin{minipage}{120mm}
\scriptsize
\begin{flushleft}
\caption[]{Input parameters}
\begin{tabular}{l|lllllll}
\hline
\hline
Model	&$\rho_{core,gas}$ & initial &$\epsilon_{SF}$ 	& T  & M$_{DM}$\\
        &($10^{-25}\rm g\, cm^{-3}$) & profile&         &(K) & $10^{11}M_{\odot}$\\
\hline
Massive ellipticals (Paper I) \\
\hline
Ma1  &      0.6            &IS         &1            &$10^6$   &22 \\
Ma2  &          0.6          &IS        &10           &$10^4$  &22  \\
Ma3  &          0.6         &IS           &2         &$10^4$   &22\\
Mb1          &  0.06       &flat           &1          &$10^7$ &22   	\\
Mb2  &        0.2          &flat         &1           &$10^5$  &22  	\\
Mb3          &  0.06       &flat            &10        &$10^6$ &22	\\
Mb4          &  0.6        &flat          &1          &$10^6$  &22   	\\
La   &         0.6         &IS         &10         & $10^7$    &57	\\
Lb   &        0.6          &flat      &10            &$10^6$  &57	     	\\
\hline
Low mass ellipticals \\
\hline											% classificazione originale
E1a & 0.3 & flat & 0.5 & $10^5$	&2\\% Sa/jnew
E1b & 0.3 & IS & 0.5 & $10^5$	&2\\ 	% Sb1/jbis
E1c & 0.3 & IS & 3 & $10^5$	&2\\	% Sb2/jter	

E2a & 0.01 &flat &$10^5$	&2\\ % S1
E2b & 0.03 &flat &1 &$10^5$ &2\\	% S1bis
E2c & 0.02 &flat &1 &$10^5$	&2\\	% S2
E2d & 0.02 & flat &0.1 &$10^5$	&2\\	% S3

E3a &0.02 &flat &0.3 &$10^5$	&2\\	% S3te
E3b &0.02 &flat &0.2 &$10^5$	&2\\	% S3qt
E3c &0.02 &flat &0.3 &$10^6$	&2\\	% S3ten
E3d &0.02 &flat &0.2 &$10^6$	&2\\	% S3qtn

E4a & 0.02 &flat &1 &$10^6$	&2\\	% S4
E4b & 0.007 &flat &1 &$10^5$	&2\\	% S41
E4c & 0.007 &flat &0.1 &$10^5$	&2\\	% S42

E5 & 0.02 &flat &3 &$10^5$	&2\\	% S5
E6 & 0.02 &flat &10 &$10^5$	&2\\	% S6
\\
\hline
Bulges \\
\hline
bulge1 & 0.02 & IS &1 &$10^6$	&20\\	% b1
bulge2 & 6. & IS &3 &$10^5$	&20\\	% b5
bulge3 & 6. & IS &3 &$10^6$	&20\\	% b6 
bulge4 & 0.02 &flat &3 &$10^5$	&20\\	% b7
bulge5 & 0.007 &flat &3 &$10^5$	&20\\	% b8
%\hline
\label{table1}
\end{tabular}
\end{flushleft}
Models called {\it E} are low-mass ellipticals, whereas models called  
{\it bulge} are
spiral bulges. The flags \emph{flat} and
\emph{IS} pertain to the initial gas distribution which can be either  
constant with radius (flat) or an
isothermal sphere (IS), respectively. The model \emph{bulge3} has  
been used in Paper II for a calibration
on the chemical properties of the resolved stars in the Milky Way bulge.
\end{minipage}
\end{table*}

We choose $R_{eff,*}$ as the radius that contains 1/2 of the stellar  
mass and, therefore, is directly
comparable with the observed effective radius, whereas we will refer  
to $R_{core,*}$ as
the radius encompassing 1/10 of the galactic stellar mass.
We did not fix $R_{core,*} = 0.1 R_{eff,*}$ a priori, in order to have
a more meaningful quantity, which may carry information on the actual
simulated stellar profile.
In most cases, this radius will correspond to $\sim 0.05 - 0.2 R_ 
{eff,*}$,
which is the typical size of the aperture used in many observational
works to measure the abundances in the innermost regions of ellipticals.

Finally, we did use the following
notation for the metallicity gradients in stars
$\Delta_{O/Fe}=([<O/Fe>]_{core}-[<O/Fe>]_{eff})/log
(R_{core,*}/R_{eff,*})$; a similar expression applies for both the
[$<Fe/H>$] and the [$<Z/H>$] ratios.
Hence, the slope is calculated by a linear regression between the  
core and
the half-mass radius, unless otherwise stated. Clearly, deviations
from linearity can affect the actual slope at intermediate radii (see  
Fig. 3).

For all the models the velocity dispersion $\sigma$ is evaluated
from the relation $M=4.65\cdot 10^5\, (\sigma/{\rm km s^{-1}})^2\, R_ 
{eff}/{\rm kpc}\, M_{\odot}$ (Burstein et al., 1997).
We warn the reader that we assume that our model galaxies are  
virialised objects
in order to assign them a stellar velocity dispersion
from their mass and effective radius, because we {do not}
model stellar kinematics.

\section{Results: global properties of the models}

We start the analysis of our results by briefly discussing some general
properties that hold for the entire sample of models - i.e. ellipticals
and bulges - åÊwhose relevant predicted properties
are listed in Table~\ref{table2} (including massive ellipticals from  
Paper I). In particular, we show
the final (i.e. after SF stops) values for the stellar mass and
effective radius, the åÊ[$<O/Fe>$] abundance ratio in the galactic  
centre and the
gradients in [$<O/Fe>$] and [$<Z/H>$].

The relation between [$\alpha$/Fe] and mass tracers
(see e.g. Worthey et al. 1992, Nelan et al. 2005, Thomas \& Davies,  
2006),
is satisfied, as shown in Fig.~\ref{figt}.
It is important to ensure that the models fulfill such a relation,
as it is the most severe test-bench for a galaxy formation scenario
(see Pipino et al. 2009a).
We note that the mass-metallicity relation is also satisfied, since  
our massive objects
have an average stellar metallicity which is super-solar, whereas
the simulated low mass ellipticals and bulges have solar metallicity  
at most.
More quantitatively, a linear fit to our model predictions gives
[$<$Z/H$>$]$_{core} = -1.14+0.57 log\, \sigma$ to be compared with
the relation [$<$Z/H$>$]$_{core} = -1.06+0.55 log\, \sigma$ inferred  
by Thomas
et al. (2005) within the same aperture for  
observed ellipticals.
The robustness of our predictions is supported by the  
fact that
our models obey to the above mentioned observational constraints.
This ensures that we investigate the relation between abundance and
abundance ratios gradients by means of models that are able
to reproduce the main chemical properties of the ellipticals.
Remarkably, the above-mentioned relations are in place already after  
0.5 - 1 Gyr since the beginning
of the star formation.

In the two following sections we highlight other main features
of model ellipticals and bulges, respectively.

\begin{figure}
%\epsscale{.80}
\includegraphics[width=8cm,height=6cm]{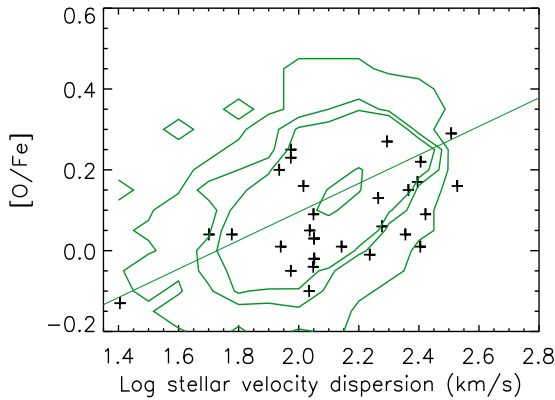}
\caption{The [$<O/Fe>$]$_{core}$-$\sigma$ relation predicted
for our model ellipticals and bulges {shown as crosses  
in the figure}.
Data from Thomas et al. (2007) are shown as contours. A liner regression
to Thomas et al. (2007) is shown by a thin solid line.
Note that in this plot we rescaled our [$<O/Fe>$] values in order to  
be consistent with
the solar abundances used by Thomas et al. (2007).
%The solid lines encompass the region constrained by observations
%(Thomas et al., 2005). In particular, we use a stripe one standard  deviation wide, centered
%around the average relation as given by Thomas et al. (2005).
Spiral bulges obey to the same relation (Thomas \& Davies, 2006).
}
\label{figt}
\end{figure}

\subsection{Elliptical galaxies}
\label{highmass}

{ In brief, we first recall from Paper I how the formation
of a galaxy proceeds in our model. We take the case La as an example.
At times earlier than 300
Myr the gas is still accumulating in the central regions where the
density increases by several orders of magnitude, with a uniform speed
across the galaxy. The temperature drops due to cooling, and the SF
can proceed at a very high rate ($\sim 10^{2-3} \rm M_{\odot}
yr^{-1}$), at variance with the outermost regions, that complete their build-up in the first 100 Myr.
This implies that a metal rich medium, dominated by SNIa ejecta, pollutes
the gas supply for the SF in the inner regions.
%We find that for\emph{flat} models (e.g. case Lb), despite
%the different initial conditions, the evolution of all the physically
%interesting quantities follows the results obtained for models \emph 
%{IS}.
After 400 Myr, the gas speed becomes positive (i.e. out-flowing gas) at
large radii, and at 500 Myr almost the entire galaxy experiences a
galactic wind. 
%The wind is supersonic for, at least, the first
%Gyr after $t_{gw}$,which is the time of the onset of the galactic
%wind and depends on the model assumptions. 
 At roughly 1.2 Gyr,
the amount of gas left inside the galaxy is below 2\% of the stellar  
mass. This gas is very hot (around 1 keV) and still flowing outside.

The galactic wind occurs {first in the outer regions and  
then in the more inner zones of the galaxy}
%externally before internally
because the work to extract the gas from the outskirts is less
than the work to extract the gas from the center of the galaxy.}
{ The age differences between internal and external zones,  
however, are less than 1 Gyr
and this ensures that our models are globally $\alpha$-enhanced. In  
this way
our models are consistent with the observed age gradients (references  
in Sec. 2) and with the [$\alpha$/Fe]-mass relation.}
The picture sketched above applies to the lower mass models
presented here.
The fact that in our galaxy formation scenario the metallicity  
gradients arise because of the different times of occurrence of galactic
winds in different galactic regions implies that the stellar  
metallicity is a function
of the local escape velocity $v_{esc}$ for all the galaxies.
In fact, in the regions where $v_{esc}$ is low (i.e. where the
local potential is weaker), the galactic wind develops
earlier and the gas is less processed than
in the regions where $v_{esc}$ is higher (see Martinelli et al. 1998).
Such a relation as been originally suggested by several authors (e.g.  
Davies et al., 1993, Peletier et al., 1990)
and now confirmed by Scott et al. (2009).
Here we can also show that the \emph{local} index-$v_{esc}$
trend matches the \emph{global} scaling (Scott et al., 2009).
{ In particular, we make use of the definition $v_{esc}=\sqrt{(-2 \Phi)}$,  where $\Phi$ is the potential
due to stars and DM,  in agreement with the definition used by the observers.
Some important caveats apply to this comparison. In observations, $v_{esc}$ depends on the modelling of the potential. 
Moreover, our models are spherically symmetric, whereas observed galaxies are not.}
In Fig.~\ref{fig0b} we show that
metallicity (given by the index $Mg_b$) versus $v_{esc}$ gradient slope for our models.
The central $Mg_b$ value for each model galaxy is given by an  
asterisk, whereas the value at 1 $R_{eff}$ by a cross. Each couple of points connected  
by a line represents a galaxy: this is the local relation.
The dashed line is the observational global (i.e. the fit to the central values
of $Mg_b$ and $v_{esc}$ in observed galaxies) trend
reported by Scott et al. (2009) along with the 3$\sigma$ dispersion  
(dotted lines). We show that the models presented in this paper reproduce the  
observed trend within the observed scatter. 
The fact that the each galactic region follows the global trend
strongly suggests the idea
that a uniform process - like the monolithic collapse - is behind
the formation of the gradients.

\begin{figure}
%\epsscale{.80}
\includegraphics[width=8cm,height=6cm]{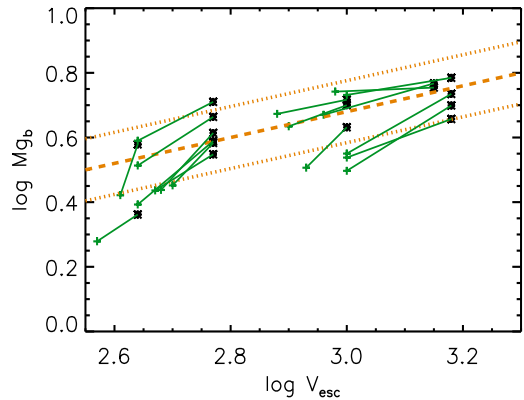}
\caption{Metallicity versus $v_{esc}$ gradient slope for our models.
The central $Mg_b$ value for each model galaxy is given by an  
asterisk, whereas
the value at 1 $R_{eff}$ by a cross. Each couple of points connected  
by a line
represents a galaxy (the local relation). The dashed line is the global relation
(Scott et al., 2009) along with the 3$\sigma$ dispersion  
(dotted lines).}
\label{fig0b}
\end{figure}

%To be more quantitative, we find metallicity ([$<Z/H>$]) gradient  
%slopes in the range -0.5 -- -0.2 dex per decade
%in radius, in agreement with the observations (c.f. Fig.~\ref{fig1}).
%As for the mass-metallicity relation, the build-up of the gradients  
%is very fast and we predict
%negligible evolution after the first 0.5 - 1 Gyr.
%We will discuss these results in {more} detail below.
%We recall that in Paper I we found that the \emph{actual} (i.e.
%mass-averaged) metallicity gradients can be flatter than the
%luminosity-weighted ones (i.e. the observed ones, see Paper I).

%\par
%\hbox{}
\begin{table*}
\centering
\begin{minipage}{120mm}
\scriptsize
\begin{flushleft}
\caption[]{Model results}
\begin{tabular}{l|llllllll}
\hline
\hline
Model	&$M_*$ & $R_{eff,*}$& [$<O/Fe>_{*,core}$] & $\Delta_{O/Fe}$ & $ 
\Delta_{Z/H}$ \\
&($10^{10}M_{\odot}$) &({kpc}) \\
\hline
Massive ellipticals (Paper I) \\
\hline

Ma1    &6.0&    12         &  0.29                   &  0.02           
& -0.19	\\
Ma2    &25.&    7.7       &  0.22                   &  -0.21           
& -0.52	\\
Ma3    &25.&    8.3        &  0.35                  &  -0.17          
& -0.03	\\

Mb1   &6.0 &    17        &  0.14                   &  0.09           
& -0.20	\\
Mb2   &3.0 &    8.7       &  0.33                  &  0.          &  
-0.18	\\
Mb3   &21&      8.8       &  0.17              &  -0.08        &   
-0.34	\\

Mb4   &26 &     5.4       &  0.42                    &   
-0.08          & -0.20	\\

La    &26&     29         &  0.14                   &  0.19            
& -0.50\\
Lb   &29&      21         &  0.12                   &  0.32            
& -0.30	\\

\hline
Low mass ellipticals \\
\hline

E1a åÊåÊ& 0.74 åÊåÊåÊåÊåÊ& 1.7	& åÊ0.08  
åÊåÊåÊåÊåÊåÊåÊåÊåÊåÊåÊåÊåÊåÊåÊåÊåÊåÊ 
åÊåÊåÊåÊåÊåÊåÊåÊåÊåÊåÊåÊ& åÊ-0.04  
åÊåÊåÊåÊåÊåÊåÊåÊåÊ& -0.26		\\
E1b åÊåÊ& 0.74 åÊåÊåÊåÊåÊåÊ& 1.7 åÊåÊ& åÊ0.36  
åÊåÊåÊåÊåÊåÊåÊåÊåÊåÊåÊåÊåÊåÊåÊåÊåÊåÊ 
åÊåÊåÊåÊåÊåÊåÊåÊåÊåÊåÊ& åÊ-0.13  
åÊåÊåÊåÊåÊåÊåÊåÊåÊ& -0.21	\\
E1c åÊåÊåÊåÊ& 0.74 & 1.7	& 0.28  
åÊåÊåÊåÊåÊåÊåÊåÊåÊåÊåÊåÊåÊåÊåÊåÊåÊåÊ 
åÊåÊåÊåÊåÊåÊåÊåÊåÊåÊåÊåÊåÊ& åÊ-0.11  
åÊåÊåÊåÊåÊåÊåÊåÊåÊ& -0.21	\\
E2a åÊåÊåÊåÊ&	1.5 &	0.9&	0.19  
åÊåÊåÊåÊåÊåÊåÊåÊåÊåÊåÊåÊåÊåÊåÊåÊåÊåÊ 
åÊåÊåÊåÊåÊåÊåÊåÊåÊåÊåÊ&-0.03&	-0.29 \\
E2b åÊ&	1.8&	0.6&	0.14 & åÊåÊåÊåÊåÊåÊåÊ+0.01&	-0.29 \\
E2c åÊåÊåÊåÊåÊ&	1.4 &	0.89 &	0.26 & åÊåÊåÊ-0.04 &  
åÊåÊ-0.30 \\
E2d åÊåÊåÊåÊåÊ& åÊåÊåÊåÊåÊ0.27 &	2.3 åÊ&	0.17  
åÊ& åÊåÊ+0.01	&-0.29\\
E3a åÊåÊ&0.88	&1.6	&0.18 åÊåÊ& -0.005 &-0.27 åÊåÊ\\
E3b åÊåÊ&0.65	&1.1	&0.11 åÊåÊ& +0.07	åÊ&-0.28 åÊåÊåÊ\\
E3c åÊ& 0.93	&1.6	&0.09 åÊåÊ& -0.01	&-0.32 åÊåÊåÊ\\
E3d åÊ& 0.6	&1.1	&0.03 åÊåÊ& +0.06	&-0.25 åÊåÊåÊ\\
E4a åÊåÊåÊåÊ&	1 åÊåÊåÊ&	1.7 &	0.16 åÊ& åÊåÊ-0.21  
åÊåÊ& -0.33 \\
E4b åÊåÊåÊåÊ&	0.35 &	0.6 &	0.22 åÊåÊ& åÊ-0.04  
åÊåÊåÊ&-0.36 \\
E4c åÊåÊ	&0.05	&0.5	&0.17 åÊåÊ& -0.01	&-0.22	\\
E5 åÊåÊåÊ	&1 åÊåÊåÊåÊ&	1.7 &	0.16 åÊåÊ& åÊ-0.20  
åÊåÊåÊ&-0.38 \\ 
E6 åÊåÊåÊ	&1	&1.7 åÊåÊåÊåÊåÊåÊ&	0.11 åÊ&  
åÊåÊ-0.16 åÊåÊåÊ&-0.34 åÊ\\
\hline
Bulges \\
\hline
bulge1 åÊ&	0.06 åÊ& åÊ2&	0. 		&0.09 åÊåÊ& -0.36 åÊåÊåÊ\\
bulge2 &	1.8 & åÊåÊåÊ1 åÊ&	0.40 åÊåÊåÊ	& -0.07 &  
åÊåÊ-0.22 åÊ\\
bulge3 &	2.3 & åÊåÊ0.8 &	0.3		&0.07 åÊåÊ& -0.36 åÊåÊåÊ\\
bulge4 &	3.7 & åÊåÊ0.7 &	0.29		&0.00 åÊåÊ& -0.37 åÊåÊåÊ\\
bulge5 &	1.0 & åÊåÊ0.4 &	0.28		&0.00 åÊåÊ& -0.30 åÊåÊåÊ\\
\hline
\label{table2}
\end{tabular}
\end{flushleft}
Models called {\it E} are low-mass ellipticals, whereas models called  
{\it bulge} are
spiral bulges.
Values predicted after the SF has finished.
\end{minipage}
\end{table*}

%\begin{figure}
%\includegraphics[width=8cm,height=8cm]{mappete.eps}  %{ofe_b6.eps}
%\caption{Solid line: [O/Fe] vs. [Fe/H] in the gas of model (mass-weighted values
%on the gridpoints of each region).
%Symbols: recent data compiled by Ballero et al. (2007).}
%\label{fig0}
%\end{figure}

\begin{figure}
\includegraphics[width=8cm,height=8cm]{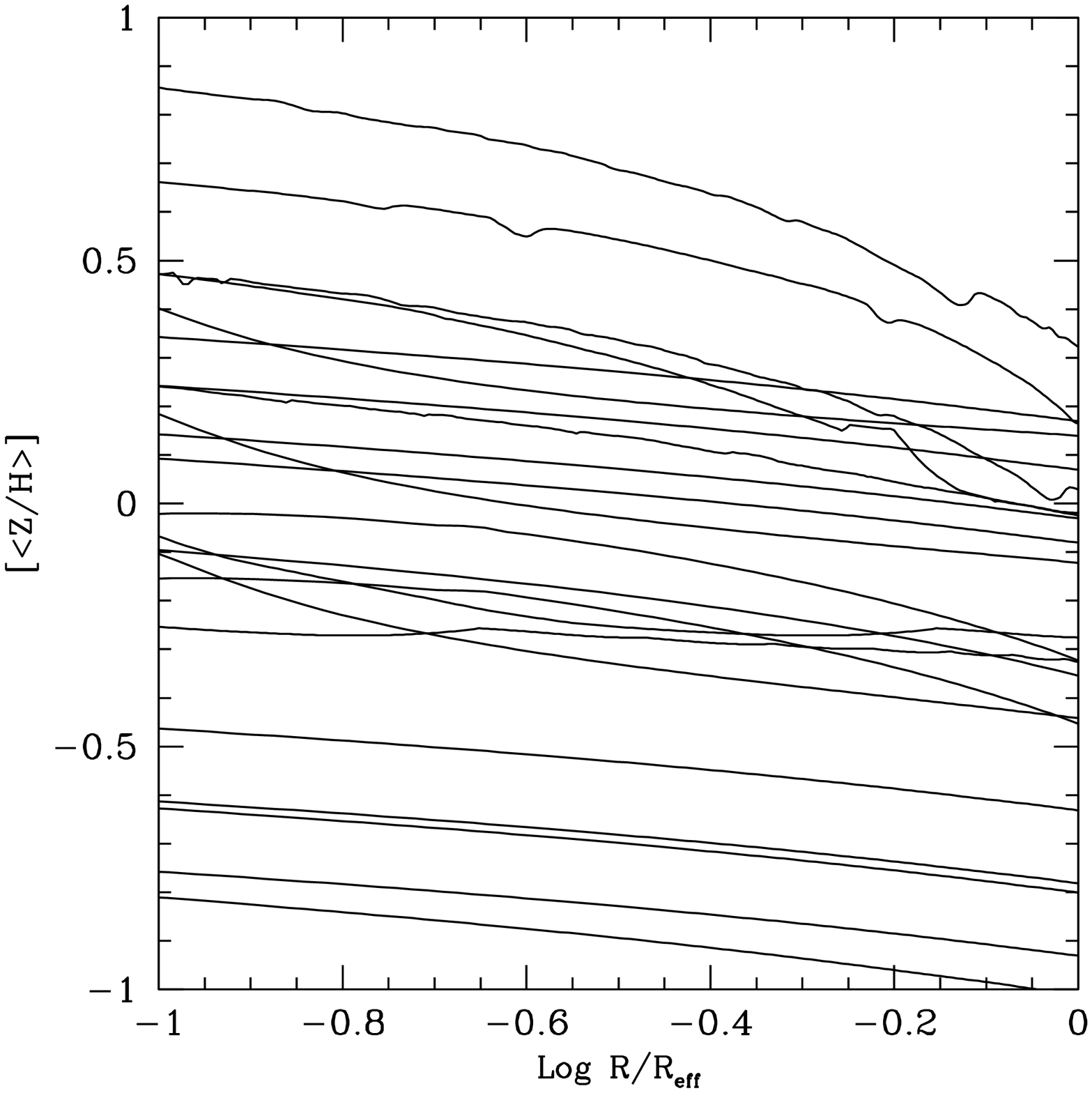}  %{ofe_b6.eps}
\caption{Metallicity profiles predicted by our model ellipticals.}
\label{prof}
\end{figure}

\subsection{Galaxy bulges}
\label{bulges}

Remarkably, all the results discussed in the previous sections apply
to smaller objects (but embedded in much more massive haloes) such as
the galaxy bulges, although the gradient slopes are slightly smaller
(see entries in Table~\ref{table2}). The main difference is that,
due to their host galaxy potential well, strong and long lasting
winds do not develop.
%For instance, our models suggest that also in spiral bulges, the  
%possible gradient in the [$<\alpha$/Fe$>$]
%ratio is related to the interplay between the velocity of the
%$\alpha$-enhanced radial flows, moving from the outer to the inner  
%galactic
%regions, and the intensity and therefore duration of the SF formation  
%process
%at any radius. 
{ We also find that the bulge formation is fast in  
agreement with the original
suggestion by Matteucci \& Brocato (1990), Elmegreen (1999) and the more recent { work} by Ballero et al. (2007).

We take advantage of the classical bulges as
a further tool to calibrate our models. Indeed, in Paper II (where we refer
the reader for further details) we compared our
model predictions to the properties of the \emph{resolved stellar  
population} observed in the Milky Way bulge by using the model
\emph{bulge3} and found a remarkable agreement. This model has a stellar mass of $\sim 2\times 10^{10}
M_{\odot}$  and a radius of $\sim 1$kpc
in order to match the observed properties of our own Galaxy bulge  
{(e.g. Minniti \& Zoccali, 2008)}. 
%In particular, 
%in fig.~\ref{fig0}
%we compare the model predictions to the observations 
%in the [$O/Fe$]-[$Fe/H$] plane.
%In particular, the solid line is the predicted average trend (which reflects the mean
%composition of the gas), while the symbols refer to
%a compilation of observed abundances in single stars of the Milky Way  
%bulge we find a remarkable agreement.
%In plotting the solid line we averaged over all the stars. These are
%born in different points of the grid, and hence may have undergone
%different chemical evolution histories.
%In Paper II we show also that 
%This explains why the lowest probability contour (i.e. the outermost)
%is quite broad.
%In the single degenerate scenario, the minimum timescale for having a 5SNIa is 30 Myr, therefore
%we can have stars forming in a cell in which [O/Fe] $<1$ even at [Fe/ 
%H]=-1.5, but this is very unlikely.
%In fact,
%Remarkably, 
%the majority of the observed data points lie within the  
%highest probability contour.
%and, at the same time, 
{ The same model reproduces the chemical  
constraints coming from the Bulge integrated light,
in that it predicts the following values for the indices H$_{\beta}$=1.61, $Mg_2$=0.29
and $<Fe>$=2.46 in good agreement with the observed values of
H$_{\beta}$=1.5$\pm$0.6, $Mg_2$=0.23$\pm$0.04
and $<Fe>$=2.15$\pm$0.4 (Puzia et al., 2002).}
This is an important point that must be stressed: the abundance
(and abundance gradients) that may be inferred from the analysis of  
Lick indices are average values. With resolved stellar populations (Paper II)
is possible to show that the models  
presented here not only explains the average values, { namely
the mean properties of a Composite Stellar Population (CSP)}, but also their evolution in the [$O/Fe$]-[$Fe/H$] plane, { namely
the composition of each SSPs that make a CSP.}
%as well as the intrinsic scatter  
%in the stellar metallicity distributions. 
}
%\emph{The minimum time is in general different for the timescale for  SNIa to be effective}
%which is tipically inferred from the position of the knee in the [O/Fe] vs [Fe/H] curve
%(dashed line in fig.~\ref{fig0}, right panel).
Our fiducial model assumes Salpeter (1955) IMF, which successfully
reproduces the properties of massive spheroids.
In Paper II åÊwe show that the stellar metallicity
distribution predicted by such a model reproduces the observed K-giant
metallicity distributions for the Milky-Way bulge. We refer to Paper  
II (c.f. Fig. 2) for the
test of other possible IMFs, motivated by either observations
or theoretical efforts, which seems more appropriate for bulges.

The reader should note that we present several other models for  
bulges which do not necessarily have
properties - such as stellar mass or radius - similar to those of the  
Milky
Way bulge.
%Nonetheless, the [$O/Fe$]-[$Fe/H$]
%and the K-giant diagrams presented in Paper II should be considered
%prototypical, therefore we do not repeat the analysis here.

\section{The predicted gradients}

\label{results}
\label{stats}

In this section we turn our attention on the main topic
of this work and investigate the possible dependence of the gradient
slope - and its scatter - from either the stellar mass or some mass  
tracers.
We first focus on the actual prediction from the modeller's point of  
view, namely
gradients in abundance and abundance ratios versus mass and central
[$<$O/Fe$>$], whereas we refer to Sec. 5.2 for our model predictions  
transformed into \emph{observational}
line-strength indices. % by means of the Thomas et al. (2003) SSP library.

{ The metallicity profiles predicted by our model ellipticals over the 0.1-1$R_{eff}$
range are shown in Fig.~\ref{prof}. In the vast majority of the cases we predict
a linear decrease of the metallicity with log (r) and thus justifies
the adopted definition for $\Delta_{Z/H}$. We refer the reader
to Paper I (Fig. 8) comparison between the observed and predicted [O/Fe] profiles for some
relevant cases.}

For elliptical galaxies we make use of Mehlert et al.'s (2003) and  
Annibali et al.'s (2007) datasets,
whose samples are larger than Ogando et al.'s one, although the  
former do not
find such a strong correlation between gradient slope and mass as the  
latter (other works with less galaxies are not taken into
account in order not to have a poor statistics). For bulges we adopt  
the data from Jablonka et al. (2007),
who explore a range in velocity dispersions similar to the above- 
mentioned articles. Unfortunately, we cannot use a homogeneous set of observables to  
constrain both
the \emph{theoretical} and the \emph{observational} predictions for  
several reasons.
In the first place, in several articles
the authors do not extract the [$<$O/Fe$>$] abundance ratio gradient  
from their line-strength indices
(e.g. Ogando et al. 2005\footnote{Noticeably, they could not convert  
the indices into abundances
in several galaxies whose combination of index values fell outside  
Thomas et al. (2003) SSP libraries.
We refer the reader to Pipino et al. (2006) and Paper I for a  
detailed discussion
on the theoretical aspects of such a problem. Here we just mention  
that SSP libraries
do not cover all the possible combinations in the space [$<O/Fe>$]- 
[$<Fe/H>$]-[$<Z/H>$],
being typically built just as functions of two of them.}, Kobayashi  
\& Arimoto 1999, and Sanchez-Blazquez et al. 2006).
Secondly, in all cases the stellar
mass is not observed, whereas only the stellar velocity dispersion is  
given. Finally, several authors
rely on a different sub-set of the Lick line-strength indices to  
infer the metallicity.

\subsection{Theoretical relations with mass and  
mass tracers}

With the above mentioned caveats in mind, in Fig.~\ref{fig1} we present our  
predictions regarding the \emph{theoretical} relation
between abundance gradients and mass tracers (namely the stellar  
velocity dispersion and the central [$<O/Fe>$]). The remainder
of this section is devoted to fully describe Fig.~\ref{fig1}.

\subsubsection{Gradients in metallicity}

Let us first focus on the upper row of Fig.~\ref{fig1}: the total metallicity
gradient. In the left panel we show the [$<Z/H>$] gradient slope
in the stellar component predicted by our model for ellipticals (hollow circles)
and bulges (full dots) as a function of the stellar mass {when all  
galaxies are considered}. This is the \emph{actual} prediction
of our models. Formal linear regression fits to the entire sample of  
model galaxies (solid
line), to the galaxies with steepest gradients (dotted line ) and to  
dwarf ellipticals (dashed line) are shown.
We predict a very mild trend in mass. In the high mass region, our  
model predictions span a range in the gradient slopes
similar to the observed values. Neither our models, nor the three  
observational samples (taken togheter)
show any sign of (anti-)correlation as suggested by the  
single authors. We therefore conclude
that, in this mass range, is more appropriate to speak of an increase  
in the scatter of
the gradient slope at a fixed mass.
{If we take only the four less massive
objects we find a quite steep relation between metallicity gradient  
and galaxy mass,
parallel to the locus of the galaxies with the steepest
gradients (we call it the \emph{maximum steepness} boundary line)  
similar to the predictions of the earlier monolithic collapse models.}
This finding seem to be in qualitative agreement with the  
observational results by Spolaor et al. (2009). {
\emph{As for the points near the \emph{maximum steepness boundary},  
they always refer to the models with the highest SF efficiency at  
that given mass. } 
We note another trend, symmetric to \emph{maximum steepness boundary}  
with respect to the solid line (trend of the entire sample), in the sense that at  
the highest masses we have
also the flattest gradients. This seems to go in the direction of Ogando et  
al. (2005), Spolaor et al. (2009) and Jablonka et al. (2007) results. In  
particular, the scatter is minimum
at masses below $\sim 10^{10}M_{\odot}$. These galaxies tend to have  
neither shallow metallicity gradients
nor very steep ones.

In order to explain such findings, we first note that
the formal linear regression to our model predictions gives $\Delta_{Z/ 
H}\sim -0.04\, log \, \sigma$,
namely a value much smaller (in absolute value) than the slope of the  
mass-metallicity relation $[Z/H]_{core}\sim 0.57\, log\, \sigma$.
Therefore the relation between gradient slope and galactic mass  
cannot be explained by the Carlberg (1984)'s argument
(c.f. Introduction, see also Jablonka et al. 2007). In other words,  
the steepening of the gradient with mass is not due to the sole
increasing metallicity of the galactic core, whereas the outermost  
regions of galaxies differing
in mass keep the same value for [$<Z/H>$]$_{eff}$.
Indeed it has been shown observationally that the metallicity of the  
entire galaxy should obey
to the mass-metallicity relation (e.g. Graves et al. 2007). Such a  
relation
is satisfied by our models, for which we predict $[Z/H]_{eff}\sim 0.53 
\, log\, \sigma$. Hence
$\Delta_{Z/H} \sim [Z/H]_{eff}(\sigma)-[Z/H]_{core}(\sigma)\sim 0.53 
\, log\, \sigma - 0.57\, log\, åÊ\sigma = -0.04\, log \, \sigma$ 
\footnote{
Note that in our simulations $log(R_{core}/R_{eff})\sim -1$ in  
majority of the cases.}.
The reason for this increase in the global galaxy metallicity with  
mass is due to the
fact {that the entire} galaxies, not only their central  
cores, should form more
efficiently as their mass increases in order to comply with the \emph 
{downsizing} trend, namely the
need to have [$<\alpha$/Fe$>$] ratios greater than zero and  
positively correlated to the mass.
This request renders the average gradient slope predicted by the \emph 
{revised monolithic}
models flatter than the earlier monolithic collapse models a la Larson.
However, galaxies with steep gradients still exists (e.g. models La  
and Lb) and lie on the \emph{maximum steepness boundary}.
On average, galaxies with mass $\sim 10^{10}M_{\odot}$ feature  
gradient slopes
quite close to the \emph{maximum steepness boundary}, therefore the  
scatter is small.
At larger masses, the average gradient is nearly one half of the \emph 
{maximum steepness boundary}
value at that mass, hence allowing for more intermediate possibilities.

It is interesting to understand what are the major causes
for such a range although we have  
{ unevenly sampled the parameter space
and despite the not very high number of simulated galaxies}.
We suggest that differences in the initial conditions  
of the
protogalactic cloud(s) can reproduce the  
observed scatter.
Here we discuss their relative role.
\begin{itemize}
\item Especially at larger masses, the higher the star formation  
parameter $\epsilon_{SF}$, the steeper the gradient (see Fig.~\ref{fig_corr}).
As an example compare the model Ma2 with Ma3 (or models E6 and E2c).  
Their initial conditions are the same but for
$\epsilon_{SF}$. The metallicity gradient predicted for the former  
model (which
has a higher star formation parameter) is much steeper than the  
latter case. { At the first order, the metal production scales
with the SF rate. Therefore, the metallicity increases faster
when $\epsilon_{SF}$ is increased and all the other parameters
are held fixed. However, the SF depends also on the cooling,
that increases at higher metallicity, and the local gas density. Another
important factor in regulating the SF is the interplay between stellar feedback and local potential.
Taking togheter all these factors, in the case in which the gas is already in place, the net product is that augmenting
$\epsilon_{SF}$ leads to a faster increase (relatively speaking) in the metallicity of the central regions,
where the potential is very deep and the gas is denser 
than in the outskirts, where an higher SF rate also implies that the conditions for the wind set in earlier.}

\item The role of the chosen profile is slightly less evident.
On average, the \emph{flat} profile leads to slightly steeper  
gradients. For instance, compare
model Mb4 with Ma1 that share the same initial condition but the  
profile. Similarly,
compare model E1b to E1a. This happens because,
while in the \emph{IS} models most of the (pristine) gas is already  
in place, in the \emph{flat}
models the majority of the gas supply to build the inner regions has  
go through
the outskirts when being accreted. Hence, the sinking gas is polluted  
by metals,
leading to a faster metal enrichment of the inner regions.
However, this effect is weaker than that caused by $\epsilon_{SF}$. For  
instance
compare model Ma2 with Mb4.

\item As for the temperature, starting from a higher value implies a  
longer
time for cooling the gas and feeding the star formation process. In a  
sense,
the effect is similar to the difference between the \emph{flat} case
versus the \emph{IS} case. For instance, on the basis of the previous
point we would expect model Ma1 to exhibit a (slightly) shallower  
gradient than
the one of model Ma3. Instead it is steeper.
However, a higher initial temperature is not enough to counterbalance
the effect of a large change in $\epsilon_{SF}$ (see model Ma1 versus  
model Ma2).

\item For \emph{flat} models, the initial gas density seems to be  
relatively
unimportant (e.g. compare models E2c and E4c) in the determining the  
slope
of the metallicity gradient.

\end{itemize}
In conclusion, we do not find a parameter that {
fully governs}
the creation of the gradient, even if $\epsilon_{SF}$ seems to be quite important.
Different - but reasonable - combination of the input parameters lead
to model properties that {obey both} the overall  
properties observed in elliptical
galaxies and exhibit average metallicity gradient of -0.3 dex per  
decade in radius.
Changes in the initial conditions within the same broad formation scenario create the scatter in the  
predicted gradients at a single mass.
These changes, therefore, should not be ascribed to different pictures for the  
formation of the galaxies. They rather mimic cases in which the accretion
from the proto-galactic clouds may be faster (e.g. the \emph{IS} cases)
or proceed through cold accretion through filaments (Dekel \& Birnboim,  
2006, the \emph{flat} case).
They also show the different behaviour of models where
the star formation is favored (higher $\epsilon_{SF}$, e.g. for the  
formation
of the most massive galaxies) or disfavored (models with high initial  
temperature: the
gas is accreted in pre-existing haloes and has to cool before forming  
stars).} 

In the {middle and right panels in the upper row of Fig.~ 
\ref{fig1}} we compare our model predictions to metallicity
gradients inferred from observations (Mehlert et al 2003,
Annibali et al. 2007, Jablonka et al. 2007). Obviously the above  
discussion on the cause of the (scatter in the) metallicity gradient applies  
also to the other mass tracers ($\sigma$ and the central [$<O/Fe>$]). As
explained above, however, here we can compare our predictions with  
the values measured by the observers. We can thus show that
the predicted range as well as the average gradient slope (-0.3 dex per decade in radius)
are in agreement with observations.
We note how different observational groups infer slightly different mean
$\Delta_{Z/H}$ (e.g. compare the samples in Fig.~\ref{fig1}).
For instance in the literature average values either as low as $-0.22 
\pm0.1$ or as high as $-0.34\pm0.08$ (Brough et al., 2007)
can be found\footnote{We refer to Table 4 in Spolaor et al. (2008,  
and references therein) for a useful
comparison of the gradients in age, metallicity and $\alpha$- 
enhancement inferred by
the above mentioned observations.}. Still consistent
with each other, though. This might be {due to a}  
different combination of line-strength
indices used to infer the variation in metallicity (see the analysis  
in Sanchez-Blazquez et al. 2006).
Also differences in the SSP library used to transform indices into
abundances can create the offset. Moreover, small number statistics  
can still bias the results
as well as the fact that, even in the same sample, metallicity  
gradients are
not measured out to the same radius.
Some authors claim the difference
is caused by the environment, with field ellipticals featuring shallower gradients
on average with respect to galaxies living in higher density regions (Sanchez-Blazquez et al. 2006).
Such a suggestion might explain the offset between the Mehlert et  al. (2003) Coma cluster ellipticals
and the Annibali et al. (2007) spheroids. %However, note that the galaxies in Mehlert et al. (2003)
%have a very high average age gradients not found in any other observational works.

\clearpage
\begin{figure}
%\epsscale{.80}
%\includegraphics[width=8cm,height=8cm]{fig1.eps}
\includegraphics[width=\textwidth,height=15cm]{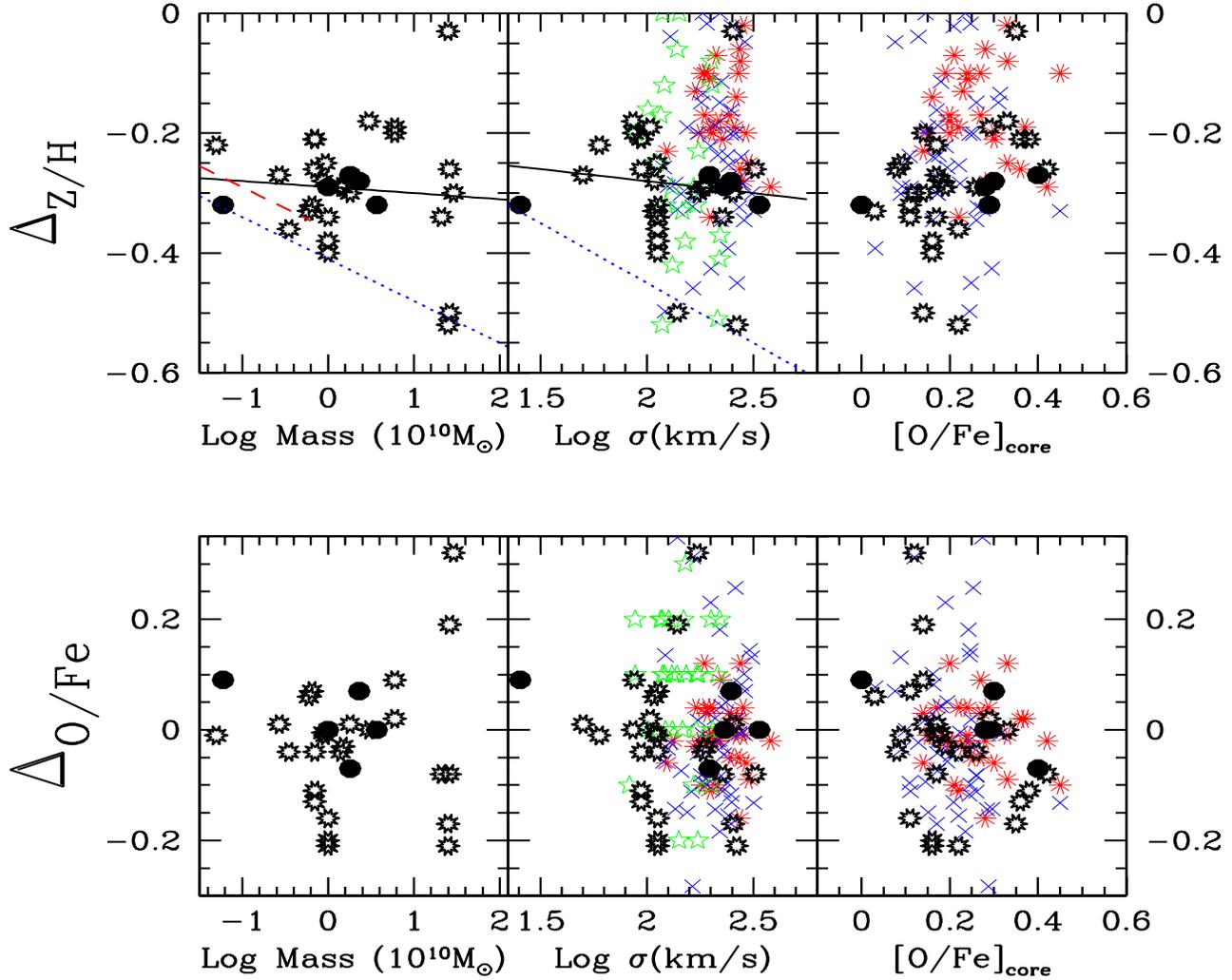}
\caption{Theoretical predictions versus abundance gradients inferred  
from observations. \emph{Upper panels}: Predicted [$<Z/H>$] gradient  
slope
in the stellar component of our model ellipticals (hollow circles) and bulges
(full dots) as a function of the stellar mass and other mass tracers
(namely the stellar velocity dispersion and the central [$<O/Fe>$]).  
Formal linear regression fits to the entire sample (solid
line), to the galaxies with steepest gradients (dotted line, the \emph 
{maximum steepness
boundary}) and to dwarf ellipticals (dashed line) are shown.
Data from Mehlert et al. (2003, asterisks, red), Annibali et al. (2007,  
crosses, blue)
and Jablonka et al. (2007, stars, green)
are shown.
\emph{Lower panels}: As above, but for the [$<O/Fe>$] gradient.}
\label{fig1}
\end{figure}
\clearpage

%Therefore, the \emph{age-metallicity} degeneracy is far from
%being solved.

\subsubsection{Gradients in abundance ratios}

We now move to the analysis of the bottom row of Fig.~\ref{fig1}.
No clear relation with mass is found for the [$<$O/Fe$>$] radial  
gradients.
%This is not a surprise,
%since our models were built to investigate the observed lack of correlation
%between [$<$O/Fe$>$] gradient slope and mass (Mehlert et al., 2003).
Indeed, as expected from Paper I and II, most of our models predict
a nearly flat [$<O/Fe>$] gradient, with some showing
either positive or negative slopes.

{ In particular, we suggest the gradient in the [$<\alpha$/Fe$>$]
ratio to be related to the interplay between the velocity of the
radial flows moving from the outer to the inner galactic
regions, and the intensity and duration of the SF formation process
at any radius.
Clearly a larger or smaller parameter of star formation can have a  
strong influence on this process.
This result implies that we do not need the merger events
in order to have a shallow [$<\alpha/Fe>$] gradient.}

In general, we find that in our models with
$\Delta_{O/Fe}\le 0$ the role of both the gas flowing inward and the  
star formation timescale increasing at large radii
is non negligible. The role of the initial temperature can be important.
If the galaxy formation process starts from hot gas (i.e. $10^{6-7}$ K),
we predict $\Delta_{O/Fe}\ge 0$ in the majority of the cases.
%: the higher
%temperature coupled with supernovae feedback slows the Fe enriched
%flows from the external radii.
%In this sense 
They are thus
similar to the quasi-monolithic chemical evolution models of Pipino \& Matteucci (2004, PM04) with not
interacting shells in which the
\emph{infall} timescale increases at shorter radii, whereas
the star formation efficiency is constant. On the other hand, models starting with cold (i.e. $10^{4-5}$ K) gas
seem to prefer a negative $\Delta_{O/Fe}$.

The sole SF efficiency seems to affect
the predicted absolute value of
the gradient slope; in fact all the models most effective in forming  
stars exhibit
the steepest slopes at the same time.
Basically, an increase in the SF efficiency enhances the differences  
between the inner core and the outskirts set by the other initial conditions.
For instance, if the gas is already in place, a high efficiency in  
forming stars boosts the outside-in process.
In such a case, the star formation process, which also locks
the metals into the stars, is fast enough in the central regions to  
avoid the contamination
of the metals flowing from larger radii. In practice
we end up in the extreme case in which the gas flows can be neglected
and $\Delta_{O/Fe}\sim 0.2$ as in the standard chemical evolution  
models (Pipino et al. 2006).
%$\rho_{gas,0}$ does not seem to play an important
%role in the build-up of the gradients.

\subsubsection{Correlations between gradients in metallicity and  
gradients in abundance ratios}

The final part of the theoretical analysis involves the study of  
possible correlations
between gradients in metallicity and gradients in abundance ratios.
As a confirmation of what said in Sec. 5.1.2,
galaxies showing the steepest positive [$<$O/Fe$>$] gradient
slopes, have also quite a strong radial decrease in the [$<$Fe/H$>$]  
ratio (Fig.~\ref{fig3}).
These galaxies are also the most massive ones.
A correlation in this sense seems to be confirmed by the
Annibali et al. (2007) data, whereas Mehlert et al.'s (2003) galaxies
exhibit values for $\Delta_{O/Fe}$ constant with $\Delta_{Fe/H}$.
%On the other hand, the models do not show any significant
%correlation in the upper panel of Fig.~\ref{fig3}.
A quantitative confirmation needs a sample statistically
richer.
Perhaps more interestingly, neither the observations nor the models
cover the region with $\Delta_{O/Fe} < 0$ and $\Delta_{Fe/H} < -0.4$:  
galaxies
with the steepest metallicity gradients undergo a strong
outside-in formation process. % a la Martinelli et al (1998).
%However, the gas dissipation is responsible for the
%higher absolute values for $\Delta_{Fe/H}$ with respect
%to standard chemical evolution models.
In galaxies with $\Delta_{O/Fe} < 0$ - namely, models that likely have a
local star formation efficiency decreasing with galactocentric radius  
- the stellar feedback is more
effective in contrasting the metal-enhanced flows; therefore,
the final $\Delta_{Fe/H}$ is smaller (in absolute value), hence  
closer to the expectations
from models which do not take into account gas flows
within the galaxy.
At the same time we predict a paucity of galaxies in the region $ 
\Delta_{O/Fe} > 0$ and $\Delta_{Fe/H} > -0.2$.
More observations are needed to confirm this suggestion.
A lack of galaxies can also be noticed in the upper left corners on  
the left panels in
Fig.~\ref{fig1}.

\begin{figure}
%\epsscale{.80}
%\includegraphics[width=8cm,height=8cm]{fig3z.eps}
\includegraphics[width=8cm,height=8cm]{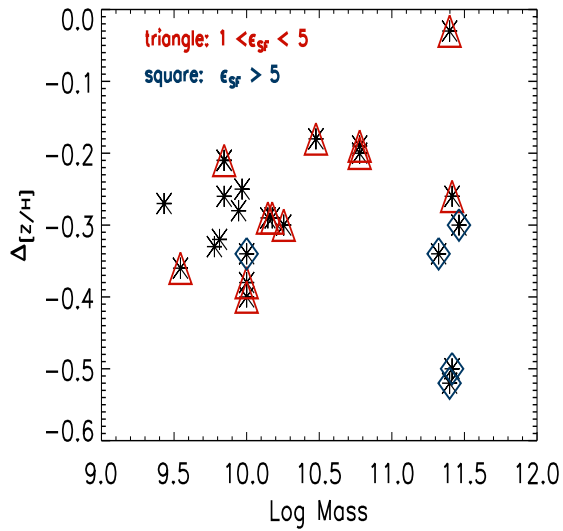}
\caption{Predicted [$<Z/H>$] gradient slope as a function of the  
mass in the case of ellipticals only (asterisks). Models are further coded by their $\epsilon_{SF}$}.
%The
%data from Mehlert et al. (2003, red), Annibali et al. (2007, blue)
%and Jablonka et al. (2007, yellow) are shown as crosses.}
\label{fig_corr}
\end{figure}

\begin{figure}
%\epsscale{.80}
%\includegraphics[width=8cm,height=8cm]{fig3z.eps}
\includegraphics[width=8cm,height=8cm]{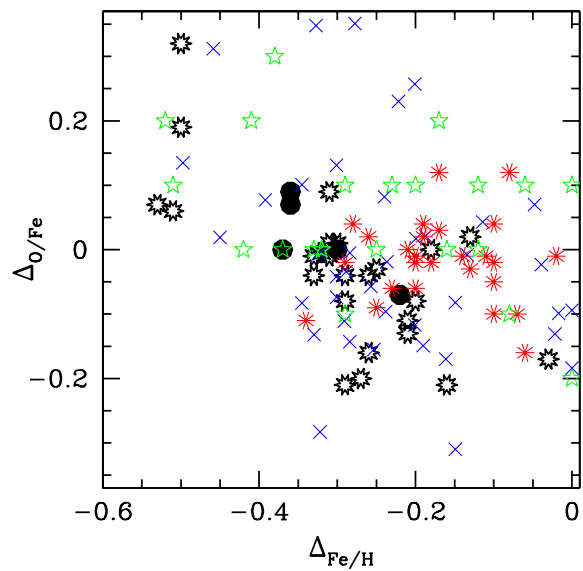}
\caption{Predicted [$<O/Fe>$] gradient slope as a function of the  
[$<Fe/H>$] one.
Symbols as in Fig.~\ref{fig1}.}
%The
%data from Mehlert et al. (2003, red), Annibali et al. (2007, blue)
%and Jablonka et al. (2007, yellow) are shown as crosses.}
\label{fig3}
\end{figure}

%On top of this, we add the warning that recent (small) episodes of SF in the center
%can further perturb the gradients, by yielding both smaller ages and average abundance
%ratios in the stars biased by the youngest SSPs (Serra \& Trager,  2007). Therefore, if they occur they
%can contribute to some degree to the scatter in the observed slopes.
%%%Finally, we also predict a flattening of the gradients at $R \le
%%%R_{core,*}$ (e.g. Fig. \ref{grad_evol}), but for the very first  gridpoints ,
%%%which are irrelevant,
%%%given the limited number of stars formed in such small regions. åÊ
%We also note that, in the region $R_{core,*}\le R \le R_{eff,*}$ the
%gradient slope is a function of the radius (see Paper I).
%%%In fact, as it can be clearly seen in Figs. \ref{grad_evol}, \ref {grad_evol2} åÊ
%%%and \ref{figura_esplicativa},
%%%in many cases the models predict a relationship between
%%%the abundance ratios and the radius which is not linear.

\subsection{Gradients in line-strength indices}

{ Before discussing the implication of our results, it is useful
to re-cast the results of Fig. 4 in terms of their observational counterparts
in order to enable a ready comparison between our model predictions and future observational
samples. This is done in Fig.~\ref{fig2}.}

We transform the predicted metallicity and abundance ratios
into line-strength indices by means of Thomas et al. (2003) SSPs. 
{ In practice, we interpolate the Thomas
et al.'s theoretical library in order to get a value for the indices Mgb
and Mg$_2$ for each combination of age, metallicity and $\alpha$-enhancement
at any given radius.}
For simplicity we assume a fixed 12 Gyr old population and
that the age radial gradient are always  
negligible,
since our models always predict age differences lower than 0.5-1 Gyr.

In the upper panels of
Fig.~\ref{fig2} we present the results for the index $Mg_2$ as a function of the stellar mass and stellar velocity
dispersion. Instead of the central [$<O/Fe>$], here we use the central
value for the index, since it correlates with mass. 
%The typical predicted gradient in the
%abundances is $d \rm Mg_2/log (R_{eff,*}/R_{core,*}) 
%\sim -0.06$ mag
%per decade in radius, again in agreement with the typical mean values
%measured for ellipticals by several authors.
The solid lines are formal linear
regression fits to the models.
The hatched area represents the portion of the
plane $\Delta Mg_2$-mass (-$\sigma$) covered by the data
from Ogando et al. (2005). 
First, we note that the trends in the
theoretical metallicity gradients with mass is confirmed when translated into  
observables.
In particular we confirm the presence of a \emph{boundary} that
corresponds very well findings by Ogando et al. (2005), whereas
the mean predicted slope ($d \rm Mg_2/log (R_{eff,*}/R_{core,*}) 
\sim -0.06$ mag
per decade in radius) is shallower than this limit at any mass.
We recall that a one-to-one correspondence between index gradient and  
metallicity gradient cannot be done, since the transformation depends also on
[$<$O/Fe$>$] and age. { This is the main reason why we based
our interpretation on Fig.~\ref{fig1} rather than on Fig.~\ref{fig2}.}
%Interestingly, the \emph{forbidden zone} is very clear in the $\Delta  
%Mg_2- Mg_2$ panel.

%In particular, null or negative gradients in the [$<$O/Fe$>$]$_{core} $ ratio, make the predicted line-strength gradients
%in low mass galaxies shallower than their \emph{theoretical}  counterparts. For some of the high mass galaxies, instead,
%the combination of steep positive gradients in the [$<$O/Fe$>$]  ratio with steep negative
%gradients in [$<$Fe/H$>$], render their gradients in the line- strength indices the steepest
%among our models.

As expected, the region of the planes index-mass tracer covered by our model  
predictions
overlaps with those of the \emph{actual} data from the observational samples (Mehlert et al  
2003 - red triangles; Annibali et al. 2007
- blue triangles; Jablonka et al. 2007, yellow; see also Sanchez- 
Blazquez et al., 2006, not shown here) { that we employed
in the previous sections.}

%were the
%correlation between gradient slope and stellar mass is much milder  
%(if any) than
%the \emph{maximum steepness boundary}.
%Our results are consistent
%with other recent \emph{revised} monolithic models; for instance,  
%Sanchez-Blazquez et al (2007)
%compared their observational results with models available in the  
%literature
%and found that galaxies are better reproduced by dissipative collapse
%models by Kawata (2001) and Chiosi \& Carraro (2002),
%rather than by merging objects as in Bekki \& Shioya (1999).
%as filled triangles).

No trend appears when we show the predictions regarding the $Mg_b$ index
(Fig.~\ref{fig2}, lower panels). We show
this discrepancy as a warning: trends can be strongly index-dependent  
(see also Sanchez-Blazquez et al. 2006). { The same relation between a theoretical metallicity
profile and the galactic mass can lead to a different gradient-mass behaviour in the observer
plane, depending on the chosen index}. According to Jablonka et  
al. (2007), such a dichotomy between these two indices should be related to Mg is due to the C  
abundance that plays a role in the index strength. Also the role of the age gradients cannot be neglected.

\section{Discussion}

{ In this paper we showed that difference in the degrees of dissipation, in the times at which  
the galactic wind occurs
and star formation histories alone can explain the observed scatter
within a \emph{quasi-monolithic} assembly.}
At variance with other authors (Kobayashi 2004) we do not
need differing channels (i.e. ``truly monolithic'' galaxies, ``truly  
hierarchical'' galaxy and
a mixture of these two) to cover the range of observed gradient slopes.
In a companion paper (Di Matteo et al., 2009) we show, instead, that  
equal mass
dry-mergers between ellipticals systematically lower (by a factor of $ 
\sim$ 2) the slope
of the pre-existing gradient. Therefore { we argue that} if one wants to explain
the scatter observed by Ogando et al. (2005), Spolaor et al. (2009)  
and Jablonka et al. (2007)
in the gradient slopes at high masses with the effects of dry-mergers
one can accommodate only a few of such episodes, otherwise we would
observe only galaxies with flat gradients.
This is true unless there is a \emph{channel} that continuously  
provides galaxies with the
steepest gradients (i.e. -0.5 dex per decade in radius) that then can  
undergo mergers.
These ellipticals clearly cannot come from mergers, otherwise we  
would need
progenitors with slopes even steeper than the early monolithic  
collapse models of Larson and Carlberg (i.e. -0.5 to -1 dex
per decade in radius, see Introduction).
Hence the majority of ellipticals have { presumably} formed in
a monolithic fashion even if we allow some dry-mergers to occur.
Therefore we refer to dry mergers as possible (but not necessary)  
episodes in the galaxy
life which åÊmay change the gradient rather than to a well-defined  
channel for galaxy
formation which co-exists with the monolithic channel as in Kobayashi  
(2004).
Similar constraints on the number of dry mergers can be obtained by  
the [$<\alpha$/Fe$>$]-mass
relation (Pipino \& Matteucci, 2008).

{ As far as the wet mergers\footnote{Wet mergers, as opposed to dry mergers, are those where gas
is involved and SF triggered.} are concerned, it is argued that
they may steepen the gradients if star formation takes
place in the metal rich gas funneled towards the galactic
core (Hopkins et al., 2008). Such a mechanism,
however, creates strong features in the age profiles - at variance with observations -
that may disappear only after several Gyr if the galaxy evolves in isolation since then,
at the expenses of a flattening of the metallicity gradient.
Unfortunately, Hopkins et al's simulation are not done in a cosmological context and start
from very simplistic assumptions (zero metallicity disks), therefore it is not 
clear what happens when the simulated galaxy undergoes several mergers as predicted
in the hierarchical formation scenario. { According to Kobayashi (2004), the steepening
of gradients by the secondary star formation (i.e., wet merger) seems to
occur only rarely.} Moreover, we expect these gradients to be erased
by subsequent dry mergers. Furthermore, it has been shown (Pipino et al., 2009a) that
the hierarchical formation is still incompatible with the observed [$\alpha$/Fe]-mass
relation in ellipticals. On the other hand, a} 
clear forecast of our model is non-evolving metallicity  
gradients in time, apart from the effects of the passive luminosity evolution  
on the galaxy spectrum.

Another prediction of our models is the correlation between [$<$O/Fe$> 
$] and [$<$Fe/H$>$] gradients,
in the sense that we expect galaxies with the steepest [$<$Fe/H$>$]  
gradients
to have a very low [$<$O/Fe$>$] abundance ratio in the core;  
therefore such
galaxies must exhibit a very steep and positive [$<$O/Fe$>$] gradient.
This does not translate into a clear correlation between the [$<$O/Fe 
$>$] gradient
and the mass because of the large scatter which erase any clear signal.

{ Before concluding, we wish to discuss some assumptions, limitations and implications of the present study.

The initial conditions where chosen in order to reproduce the typical present-day colours, SF rates
(as observed in high redshift progenitors) and central  [$<$O/Fe$>$], [$<$Z/H$>$] values 
for elliptical galaxies. Preliminary exploration of the parameter space led
us to restrict our analysis to a range 0.1-10 in the star formation parameter $\epsilon_{SF}$,
10$^{4-7}$K in the initial temperature, 10\% SN efficiency. { The majority of} the resulting models show 
metallicity profiles linearly decreasing with log radius, stellar mass to light ratios
and radii in agreement with observations. 

As a first approximation, our bulge models do not take into account the presence
of a disk. This is justified by the following two reasons: the disk forms
on a much longer timescale (e.g. Zoccali et al., 2006 Matteucci \& Brocato, 1990, Ballero et al.,2007) and current observational samples have been derived by accurately
selecting edge-on galaxies where the contamination from the discs should be minimal.
However we stress that our results for the bulges might be less constrained and robust that those concerning elliptical galaxies and further investigation is needed.

The number of modelled galaxies is small, and hence can suffer from the same small number statistics
that bias the observations. We therefore avoided any specific prediction
on the mean trend of both the metallicity and the [$<$O/Fe$>$] gradient with mass. The formal
linear regressions shown in the figures are for the mere purpose of guiding the eye and
the exact positioning of the \emph{maximum steepness} boundary might depend on the portion of the parameter
space explored. 
We stress that the main purpose of the paper is to show that even in the monolithic
formation collapse a range in the predicted gradients consistent with observations must be expected
and that a typical metallicity decrease of 0.2 dex out to 1 $R_{eff}$ can be easily reproduced
by recent monolithic collapse models. This result is robust, because even if the models presented
here do not cover the entire parameter space,  the range in predicted gradients cannot 
be decreased by adding other models. The \emph{statistical scatter} may change; however
observational samples likely suffer from the same small number statistics problem: this is why we avoided
any detailed statistical analysis in the present study.
The presence of elliptical galaxies with flatter gradients than the original models
by Carlberg (1984) should not be used as an evidence for mergers.

{ The metallicity profiles of some of the galaxies with the steepest gradients slightly depart from linearity (c.f. Fig. 3).
We would predict a shallower gradient than the one reported in Table 2 if we limited our analysis to the inner ~1/3 $R_{eff}$; vice-versa
the predicted gradient would be steeper if we were to consider only the region ~1/3 - 1$R_{eff}$. 
Therefore, we caution the reader that the conclusions about the steepest gradients in our model galaxies
and their relation with the monolithic boundary might depend on the chosen radius.
A detailed
comparison between model profiles and single well studied galaxies over a large mass range
will allow us to study the metallicity gradients in their finer details and better constrain the models presented here.

Moreover, while most of our models obey to the mass-size relation for ellipticals (e.g. Shankar et al., 2010),
some galaxies with similar mass (e.g. compare models Ma2 and La) have quite different radii.
The former model has a radius consistent with those for normal ellipticals of that mass (e.g. Shankar et al. 2010),
the latter is more typical of an early type Brightest Cluster Galaxy (BCG, e.g. Graham et al., 1996). We chose
not to make any distinction between BCGs and normal ellipticals in our models since
gradients measured in BCGs have traditionally been included in the sample as the ones that we use
and because there is no difference as far as the chemical properties are concerned (e.g. von der Linden
et al., 2007,  Brough et al., 2007).
However the reader should keep in mind that a structural difference between BCGs and normal ellipticals
seems to exist, and BCGs seem to harbor steep gradients (e.g. Brough et al., 2007). Therefore, in light of the special
role of BCGs (e.g. Pipino et al., 2009b, von der Linden et al., 2007, and references therein),
further and dedicated observations and modelling are required to ascertain if there is
any systematic difference in the metallicity gradients with respect to more ordinary ellipticals
and what is the cause.}

Also, we remind that the majority of the observational works
use Mg as a proxy for the $\alpha$ elements, as can be easily
observed in absorption in the optical bands giving rise to the well
known $Mg_2$ and $Mg_b$ Lick indices. However,
the state-of-the-art SSPs libraries (Thomas et al. 2003, Lee
\& Worthey, 2009) are computed as functions of the 
\emph{total} $\alpha$-enhancement
and of the total metallicity. This is true also for the stellar tracks,
where the O abundance dominates the opacity, and hence the stellar evolution. 
The latest observational results (Mehlert et
al. 2003, Annibali et al. 2006 and Sanchez-Blazquez et al. 2007) that we contrasted
to our predictions in this study, have been translated into theoretical ones by means of these SSPs;
therefore the above authors provide us with radial gradients in [$\alpha$/Fe], instead of [Mg/Fe].
This is why in this paper we focus on the theoretical evolution
of the $\alpha$ elements by using O that is by far the most important.

Here we briefly recall that both O and Mg come from the hydrostatic burnings in massive stars, therefore
they are produced in lock-step. It has been suggested recently (e.g. McWilliam
et al., 2007) that this might not be true at solar (and above solar) metallicities.
While this is an important effect in detailed chemical evolution studies 
it has no importance when the luminosity weighted properties of a composite stellar population
are concerned. This happens because luminosity averages weigh more the stellar populations
at lower metallicities (lower M/L), where the differences between O and Mg production
are negligible (if any). Finally, even if the abundance of O and Mg are offset
by some fixed quantity (i.e. [O/H]=[Mg/H]+const), the predicted gradient
would be the same.}

\section{Conclusions}
\label{concl}

In this paper we study the formation and evolution of
ellipticals and bulges by means of a hydrodynamical model  
(c.f. Paper I and II, respectively)
in order to understand the origin of the observed scatter in the abundance
gradients of early type galaxies.
Here we summarise our main results.

\begin{itemize}
\item We find $\Delta_{Z/H}$ in the range -0.5 -- -0.2 dex per decade
in radius with a mean value of -0.3 dex per decade in radius, in
agreement with the observations (e.g. Kobayashi \& Arimoto, 1999). åÊ

\item In agreement
with Ogando et al. (2005) and Jablonka et al. (2007), we find that 
that the scatter in the gradient slopes increases as a  
function of
mass. We reproduce such a scatter in the observations by means
of variation in the initial conditions in galaxy models.

\item Model galaxies which behave as the earlier \emph{monolithic  
collapse}
models by Larson (1974) and Carlberg (1984) define a \emph{maximum  
steepness boundary}
in the metallicity (and index) gradient slope-mass plane. \emph{These  
galaxies are preferentially those with the
highest star formation efficiency at that given mass.}

\item No galaxies with gradients steeper (i.e. more negative) than
the value given by the our predicted \emph{theoretical boundary} are  
observed
(Ogando et al., 2005, and Spolaor et al., 2009,
for ellipticals and Jablonka et al., 2007, for bulges).

\item No correlation between $\Delta_{O/Fe}$ and other galactic  
properties are
found, in agreement observations for ellipticals (Mehlert et al.,  
2003, Annibali et al., 2007)
and bulges (Jablonka et al., 2007).

\item The abundance gradients in the
abundances, once transformed into
line-strength indices, lead to
$d Mg_2/log (R_{core,*}/R_{eff,*})\sim -0.06$ mag and
$d Log Mg_b/log (R_{core,*}/R_{eff,*})\sim -0.1$ åÊ
per decade in radius, again in agreement with the typical mean values
measured for ellipticals and bulges.

\item We note that the behaviour of the gradient slope as a function of the 
galactic mass
strongly depends on the particular line-strength index used. 
In fact, the predicted $Mg_2$  
index gradient seems to correlate with mass, whereas the
$Mg_b$ index gradient does not.

\item In Paper I we demonstrated that the differential occurrence of
galactic winds (outside-in formation) alone can explain the existence  
of the metallicity gradients
discussed in this paper. Here we add that this mechanism predicts a tight  
correlation between
line-strength
index and escape velocity gradients which has been confirmed
by recent data (see Scott et al., 2009).
\end{itemize}

Larger, homogeneous and statistically meaningful observational sample of
gradients in elliptical galaxies out to one effective radius can confirm
such a prediction and validate the model.

\clearpage

\begin{figure}
%\epsscale{.80}
%\includegraphics[width=8cm,height=8cm]{fig2.eps}
\includegraphics[width=\textwidth,height=15cm]{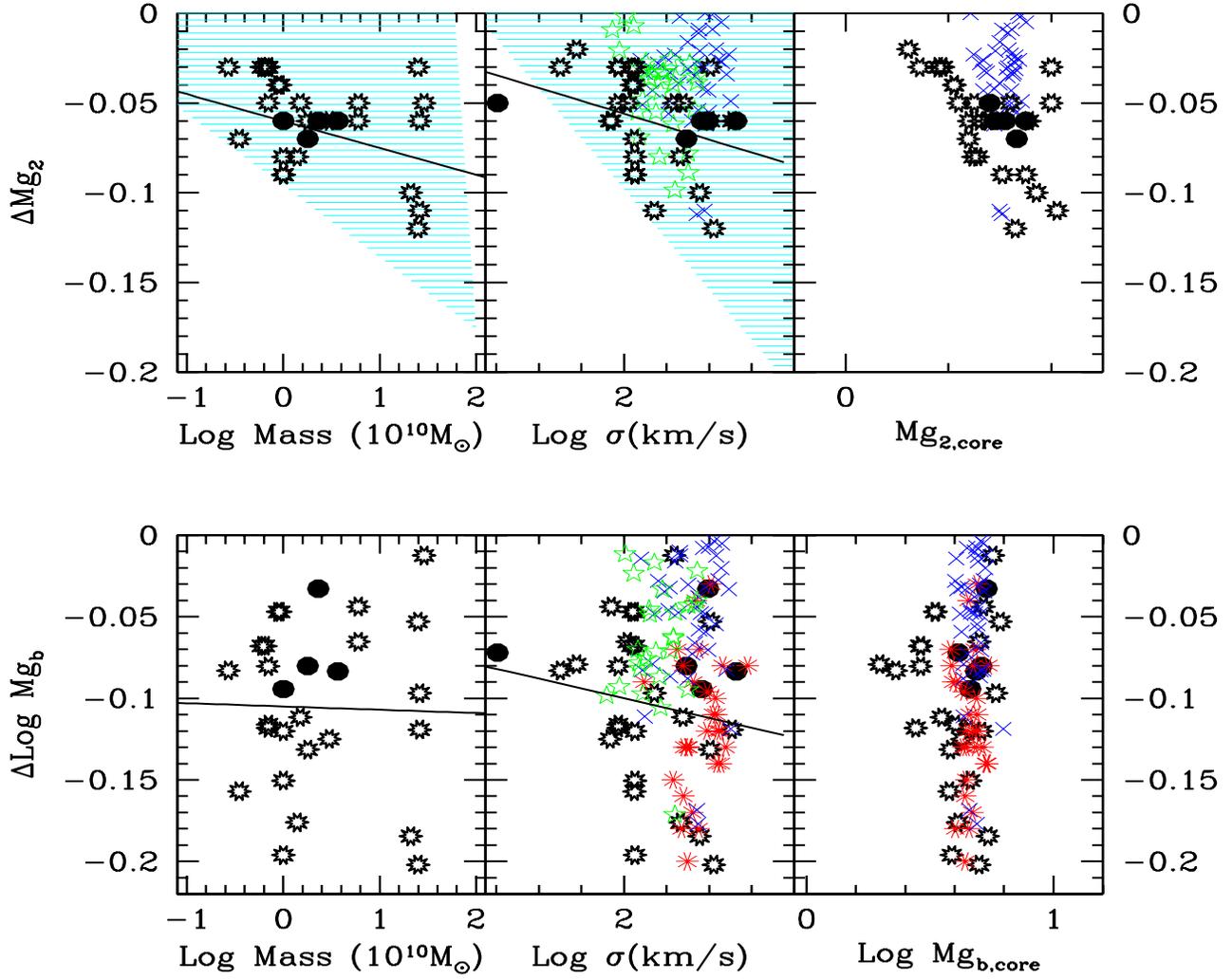}
\caption{Comparison between model predictions and data in the  
observer's plane.
\emph{Upper panels}: Predicted $Mg_2$ gradient slope
as a function of the stellar mass and mass tracers.
The hatched area represents the portion of the
plane $\Delta Mg_2$-mass (-$\sigma$) \emph{allowed} by the data
from Ogando et al. (2005). 
%Crosses:
%data from Mehlert et al. (2003, red), Annibali et al. (2007, blue)
%and Jablonka et al. (2007, yellow).
Symbols and data as in Fig.~\ref{fig1}.
\emph{Lower panels}: As in the upper panel, but for the $Mg_b$ index.}
\label{fig2}
\end{figure}

\clearpage

%\acknowledgment
\section*{Acknowledgments}
This work was partially supported by the Italian Space Agency 
through contract ASI-INAF I/016/07/0.
F.M., C.C. and A.D. acknowledge financial support from PRIN-MIUR  
2007, Prot. N.2007JJC53X .
C.C.acknowledges financial support from the Swiss National Science Foundation.
We warmly thank P. Sanchez-Blazquez, M. Spolaor, D. Forbes, S. Faber,  
P. Jablonka, M. Cappellari, N. Scott and R.Davies
for stimulating discussions. 
We thank the referee for comments that greatly improved the quality of the presentation.

\clearpage

\end{document}